\documentclass[aps,prl,twocolumn,showpacs,amsmath,amssymb,superscriptaddress,citeautoscript,longbibliography]{revtex4-1}

\usepackage{graphicx}
\usepackage{natbib}
\usepackage{dcolumn}
\usepackage{bm}
\usepackage{color}

\begin{document}

\title{Pressure-induced ferromagnetism due to an anisotropic electronic topological transition in Fe$_{\textbf{1.08}}$Te}

\author{K.\ Mydeen}
\affiliation{Max-Planck Institute for Chemical Physics of Solids, N\"{o}thnitzer Str.\ 40, 01187 Dresden}

\author{D.\ Kasinathan}
\affiliation{Max-Planck Institute for Chemical Physics of Solids, N\"{o}thnitzer Str.\ 40, 01187 Dresden}

\author{C.\ Koz}
\affiliation{Max-Planck Institute for Chemical Physics of Solids, N\"{o}thnitzer Str.\ 40, 01187 Dresden}

\author{S.\ R{\"o}{\ss}ler}
\affiliation{Max-Planck Institute for Chemical Physics of Solids, N\"{o}thnitzer Str.\ 40, 01187 Dresden}

\author{U.\ K.\ R{\"o}{\ss}ler}
\affiliation{Leibniz-Institut f\"{u}r Festk\"orper- und  Werkstoffforschung IFW, Helmholtz Str.~20, 01171 Dresden, Germany}

\author{M.\ Hanfland}
\affiliation{ESRF, BP 220, F-38043 Grenoble Cedex 9, France}

\author{A.\ A.\ Tsirlin}
\affiliation{Experimental Physics VI, Center for Electronic Correlations and Magnetism, Institute of Physics, University of Augsburg, 86135 Augsburg, Germany}

\author{U.\ Schwarz}
\affiliation{Max-Planck Institute for Chemical Physics of Solids, N\"{o}thnitzer Str.\ 40, 01187 Dresden}

\author{S.\ Wirth}
\affiliation{Max-Planck Institute for Chemical Physics of Solids, N\"{o}thnitzer Str.\ 40, 01187 Dresden}

\author{H.\ Rosner}
\affiliation{Max-Planck Institute for Chemical Physics of Solids, N\"{o}thnitzer Str.\ 40, 01187 Dresden}

\author{M.\ Nicklas}
\email{nicklas@cpfs.mpg.de}
\affiliation{Max-Planck Institute for Chemical Physics of Solids, N\"{o}thnitzer Str.\ 40, 01187 Dresden}

\begin{abstract}
A rapid and anisotropic modification of the Fermi-surface shape can be associated with abrupt changes in crystalline lattice geometry or in the magnetic state of a material. In this study we show that such an electronic topological transition is at the basis of the formation of an unusual pressure-induced tetragonal ferromagnetic phase in Fe$_{1.08}$Te. {\color{black}Around 2 GPa,} the orthorhombic and incommensurate antiferromagnetic ground-state of Fe$_{1.08}$Te is transformed {\color{black}upon increasing} pressure into a tetragonal ferromagnetic state via a conventional first-order transition. On the other hand, an isostructural transition takes place from the paramagnetic high-temperature state into the ferromagnetic phase as a rare case of a `type 0' transformation with anisotropic properties. Electronic-structure calculations in combination with electrical resistivity, magnetization, and x-ray diffraction experiments show that the electronic system of  Fe$_{1.08}$Te is instable with respect to profound topological transitions that can drive fundamental changes of the lattice anisotropy and the associated magnetic order.
\end{abstract}

\date{\today}
\maketitle


The overwhelming majority of structural phase transitions in crystalline materials is associated with changes of the symmetry or modifications of atomic positions in the unit cell. {\color{black}Only very few systems are known so far to exhibit a symmetry-conserving or `isostructural' phase transition involving pronounced
variations in the metric of the unit cell.}  
These so-called `type 0' transformations are first-order transitions for fundamental reasons \cite{Cowley_1976,Christy_1995}. The examples are driven by a diversity of mechanisms that change the internal state of the material without breaking crystalline symmetry, {\color{black}\textit{e.g.}, changes of coordination in complex framework lattices {\color{black}\cite{Saha_2013}} or electronic transitions, including valence transitions \cite{Larson_1948,Lanata13,Sarrao96}, metal-insulator transitions \cite{Akahama_1995}, or other changes of the electronic structure \cite{Rosner_2006}.}

Transformations involving spin-order with magneto-elastic couplings can appear as isostructural transitions, although time-reversal symmetry is broken. {\color{black} If a transition is driven by magnetic ordering, then  marked jumps of the lattice anisotropy are not expected in $3d$-electron systems, such as Fe$_{1+y}$Te.
However, another type of transition with magneto-elastic coupling can be envisaged which is {\em driven by a symmetry-conserving instability of the lattice}. Such a `type 0' transition mode gives rise to magnetic ordering as secondary effect.}
The resulting phase, an anisotropically deformed isostructural lattice with magnetic order as by-product, is reached  through a strongly discontinuous first-order transformation process. In metallic systems, abrupt changes in the unit-cell dimensions can be associated with modifications of the Fermi-surface (FS) topology, resulting in an electronic topological transition (ETT) \cite{Novikov_1999,Liu_2010,Glazyrin_2013,Manjon_2013,Dubrovinsky_2015}. {\color{black} We note, the changes in the FS topology discussed here are not related with topologically protected surface  states.}

In this work, a thermal transition in Fe$_{1.08}$Te under pressure is identified as an isostructural transition that is marked by a pronounced change of the axis ratio $c/a$ in its tetragonal phase. {\color{black}This type-0 transformation is driven by lowering temperature, in contrast to the vast majority of type-0 transitions known in other materials which are normaly driven by a non-thermal control parameter, such as hydrostatic pressure. Furthermore, in Fe$_{1.08}$Te the type-0 transformation triggers a ferromagnetic (FM) ordering as a secondary effect. }
Electronic structure calculations explain the transition by a drastic change of the FS topology, as suggested originally by Lifshitz \cite{Lifshitz_1960}.

{\color{black} The non-superconducting parent compound of the Fe-based superconductors, Fe$_{1+y}$Te, has prompted the interests of the condensed matter community at large \cite{Deguchi_2012}.} Bulk Fe$_{1+y}$Te exhibits a plethora of structural and magnetic phase transitions as a function of excess Fe content ($y$), temperature ($T$) \cite{Rodriguez_2011,Zaliznyak_2012,Koz_2013,Rodriguez_2013,Machida_2013,Cherian_2014}, and pressure ($p$) \cite{Okada_2009,Takahashi_2010,Koz_2012,Bendele_2013,Jorgensen_2013,Jorgensen_2015}. At low concentrations of $y\leq 0.11$, Fe$_{1+y}$Te undergoes a structural and magnetic phase transition from a high-$T$ tetragonal and paramagnetic (PM) semimetal to a low-$T$ monoclinic antiferromagnet (AFM). For $y \geq 0.12$, an orthorhombic incommensurate AFM state is realized at low temperatures \cite{Bao_2009,Rodriguez_2011,Koz_2013,Materne_2015}. Similarly, applying pressure on a sample drives the system through a series of phase transitions that closely resemble those induced by excess Fe content \cite{Koz_2012}. One such transition, being completely unanticipated, is of particular interest: a pressure-induced FM phase transition at low temperatures {\color{black} \cite{Bendele_2013,Jorgensen_2016,Fedorchenko_2011,Monni_2013,Ciechan_2014}.} This phase has not been observed at ambient pressure for samples of any level of Fe excess.

{\color{black} Details on the preparation of polycrystalline Fe$_{1.08}$Te samples by a solid-state reaction and on the electrical resistivity, magnetization, and x-ray diffraction (XRD) experiments under pressure as well as on the density-functional theory based calculations can be found in the Supplemental Material \cite{suppl}.}

\begin{figure}[b!]
\centering
\includegraphics[width=.9\linewidth]{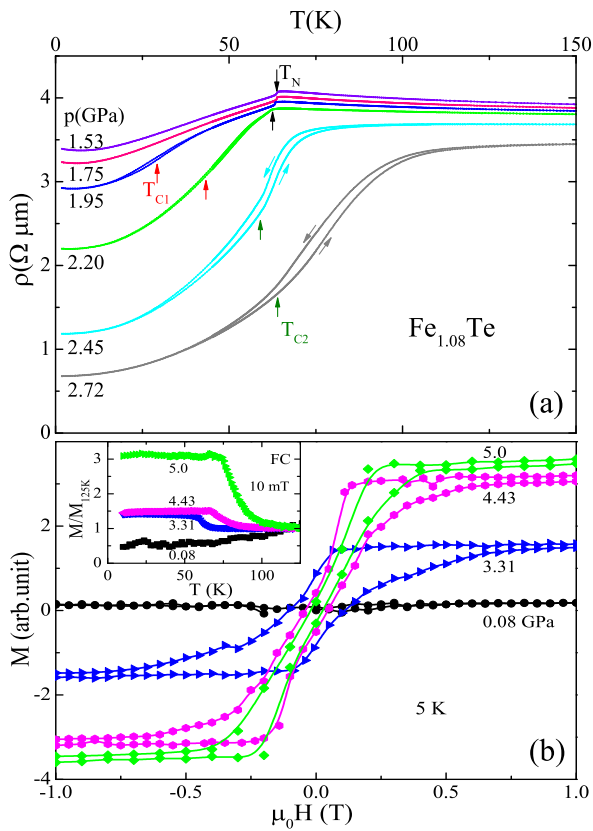}
\caption{(a) $\rho(T)$ data of Fe$_{1.08}$Te for selected pressures collected upon cooling and heating as marked by arrows. $T_N$, $T_{C1}$, and $T_{C2}$ are indicated. (b) $M(H)$ hysteresis loops of Fe$_{1.08}$Te at 5~K and at different pressures. Inset:  field-cooled $M(H)$ data for selected pressures.}
\label{data}
\end{figure}

The temperature dependence of the electrical resistivity $\rho$ of Fe$_{1.08}$Te taken in cooling and heating cycles for selected pressures $1.53 {\rm ~GPa} \leq p \leq 2.72$ GPa is displayed in Fig.\,\ref{data}a. At high temperatures, above the magnetic ordering, we observe a characteristic change in $\rho$ around 2.2~GPa. While $\rho(T)$ increases slightly upon lowering temperature at $p\leq 2.20$~GPa, indicating a semi-conducting-like behavior, $\rho(T)$ decreases at higher pressures as expected in a metal.

At 1.53 GPa, a small step-like feature in $\rho(T)$ around $T_N=65$~K indicates the phase transition from a tetragonal PM to an orthorhombic incommensurate AFM phase \cite{Jorgensen_2013}. This step-like feature is in contrast to the continuous anomaly observed at lower pressures.
It does not show any thermal hysteresis, pointing at a second-order character of the phase transition.
Upon further cooling, metallic behavior is observed down to 7~K, below which $\rho(T)$ exhibits a small upturn. The latter feature is less pronounced at higher pressures, but remains visible up to 2.20~GPa. The height of the step-like feature marking $T_{N}$ is suppressed upon increasing pressures. It is still clearly visible at 1.95~GPa, where a thermal hysteresis in $\rho(T)$, centered around 30~K and extending over a wide temperature range, is detected well below $T_{N} = 63$~K. The hysteresis is related to a first-order phase transition at $T_{C1}\approx30$~K. Measurements of $\rho(T)$ in different magnetic fields (see Supplemental Material \cite{suppl}) give a first hint that $T_{C1}$ is associated with a phase transition from the incommensurate AFM phase to a ferromagnetically ordered state at lower temperatures. A very weak anomaly in $\rho(T)$ at $T_{N}$ still exists at 2.20~GPa. Increasing pressure drives $T_{C1}(p)$ rapidly toward higher temperatures. Concurrently, the temperature interval of the thermal hysteresis decreases, but extends almost up to $T_{N}$. This observation suggests a competition between the FM and incommensurate AFM ordering.

For $p \gtrsim 2.4$~GPa, only a broad feature with a large thermal hysteresis remains in $\rho(T)$ signaling a first-order phase transition from the PM to the FM state at $T_{C2}$ [$T_{C2}$ is defined by the kink in the warming curve]. Furthermore, the resistivity decreases monotonously toward the lowest temperatures. $T_{C2}(p)$ increases strongly with increasing pressure accompanied by an increase in the width of the thermal hysteresis. The positive field dependence of the anomaly at $T_{C2}$ is consistent with a FM nature of the ordered state (see Supplemental Material \cite{suppl}).

\begin{figure}[t!]
\centering
\includegraphics[width=.9\linewidth]{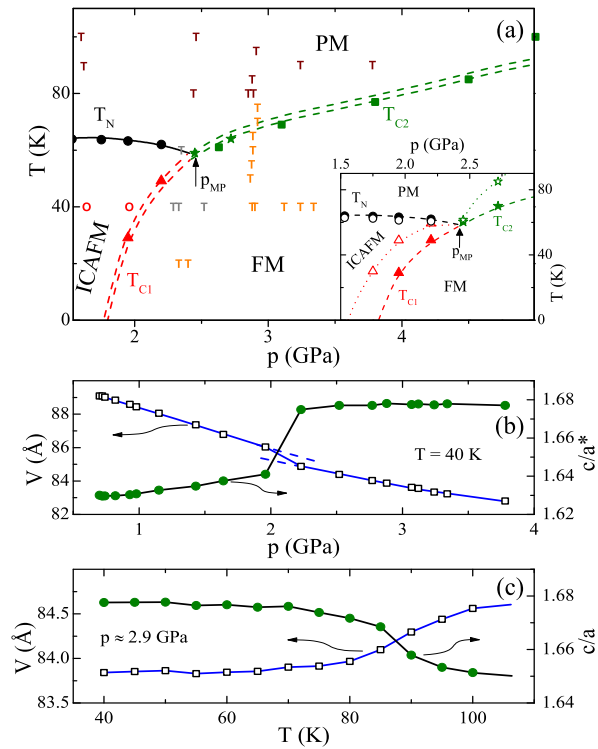}
\caption{(a) $T-p$ phase diagram for Fe$_{1.08}$Te. $\blacktriangle$, $\bigstar$, and $\bullet$ mark the transition temperatures extracted from $\rho(T)$ and $\blacksquare$ that from $M(T)$. The letters T and O represent tetragonal and orthorhombic phases according to XRD measurements. The single line and the double lines indicate second-order and first-order phase transitions, respectively. The inset displays the $T-p$ phase diagram at zero-field (solid symbols, dashed lines) and at 9~T (open symbols, dotted lines). (b) and (c) Unit-cell volume $V$ (left axis) and $c/a^*$ ratio (right axis) at 40~K as function of pressure, and at $\sim2.9$~GPa as function of temperature, respectively.}
\label{phasediagram}
\end{figure}


At 5~K, the hysteresis in the magnetization $M(H)$ loops confirms the FM nature of the high-pressure phase (see Fig.\,\ref{data}b). No signal can be resolved in the AFM phases at low pressures. \textcolor{black}{Due to the large uncertainty in the determination of the sample mass, we cannot provide absolute values of the magnetization. However, this does not affect the relative changes between different pressures. Our finding is further supported by neutron diffraction data on Fe$_{1.141}$Te, which indicate that all Fe moments are oriented along the $c$-axis in the pressure-induced FM phase \cite{Jorgensen_2016}.} The saturated high-field magnetization increases with increasing pressure, signifying a stabilization of ferromagnetism under pressure. $M(T)$ taken upon cooling in 10~mT shows a strong increase toward low temperatures upon entering the FM phase (inset of Fig.\,\ref{data}b). The extracted transition temperatures $T_{C2}(p)$ are in good agreement with the ones obtained from $\rho(T)$.


The $T-p$ phase diagram in Fig.\,\ref{phasediagram}a summarizes the results of our $\rho(T)$ and $M(T)$ investigations and, additionally, includes the XRD data taken on Fe$_{1.08}$Te. One second-order ($T_N$) and two first-order ($T_{C1}$ and $T_{C2}$) phase-transition lines meet at a multicritical point at  $p_{\rm MP} \approx 2.4$~GPa and $T_{\rm MP} \approx 58$ K. A field of 9~T suppresses $T_{N}$ only weakly, but strongly enhances $T_{C1}$ and $T_{C2}$, except for the very vicinity of the multicritical point (Fig.\,\ref{phasediagram}a, inset). The strong increase in $T_{C1}$ and $T_{C2}$ is expected for a ferromagnet. We note that while at 1.75~GPa in zero field no $T_{C1}$ anomaly in $\rho(T)$ is resolved down to 1.8~K, a field of 9~T induces FM ordering below $T_{C1}^{9{\rm T}}\approx30$~K.

By entering the FM phase from the incommensurate AFM upon increasing pressure the lattice symmetry changes from orthorhombic to tetragonal. The XRD data taken at 40~K upon increasing pressure indicate a broad two-phase region dominated by the tetragonal phase in a pressure window between 2.1 and 2.5~GPa consistent with the large hysteresis observed in the electrical resistivity. The transition from the orthorhombic-incommensurate AFM to the tetragonal-FM phase at $T_{C1}$ is accompanied by a reduction of the unit-cell volume $V$ and an increase in the ratio of the $c$ and $a^*$-axis lattice parameters [for orthorhombic and monoclinic symmetries $a^*$ = $\frac{1}{2}(a+b)$]. At $T_{C1} = 40$~K we find a marked change in both $V$ and $c/a^*$ by $-0.6\%$ and $2.1\%$, respectively, upon the transformation from the orthorhombic AFM to the tetragonal FM state (see Fig.\,\ref{phasediagram}b). This transformation  connects two phases with different and unrelated magnetic symmetries, which explains why a classical first-order-type behavior with hysteresis is observed.

At $T_{C2}$, in contrast, the {\color{black} transformation} from the high-temperature PM to the low-temperature FM state also takes place as a first-order process, evidenced by the large hysteresis observed in the $\rho(T)$ data, but without change of the tetragonal lattice symmetry. At this symmetry-conserving phase transition at $T_{C2}$, again marked variations in $V$ and $c/a$ are seen, \textit{e.g.}\ $-0.8\%$ and $1.5\%$ at $\sim2.9$~GPa, respectively (see Fig.\,\ref{phasediagram}c). This testifies that {\color{black} this phase transition is not driven by magnetic ordering}. Rather, the marked change of the lattice aspect ratio and its volume suggests that the internal electronic structure of the material is fundamentally transformed when entering the FM state. {\color{black} The reason for the transformation, therefore, is not a symmetry-breaking in the spin-structure, but an isostructural instability of the tetragonal lattice.} We note that the changes in $V$ and $c/a$ upon entering the tetragonal-FM phase are similar, independent of the path, \textit{i.e.} upon entering from the orthorhombic-incommensurate AFM or from the tetragonal-PM phase. In the following we turn to electronic-structure calculations to track the FS evolution across the symmetry-conserving phase transition.

\begin{figure}[t!]
\centering
\includegraphics[width=.9\linewidth]{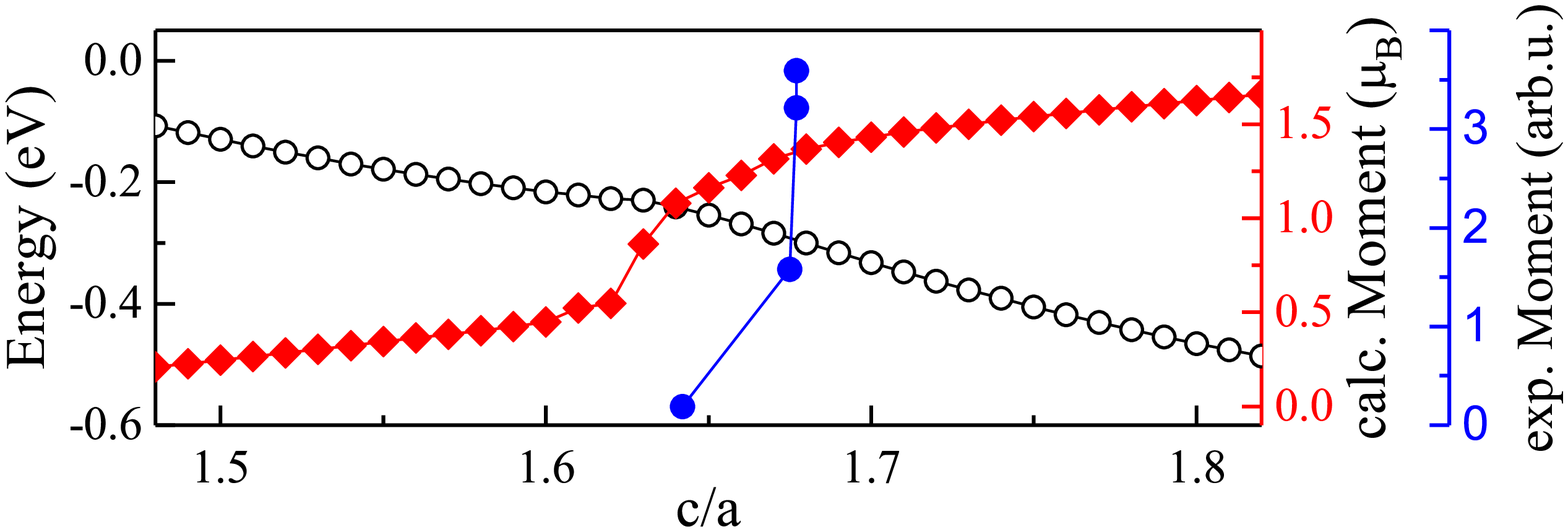}
\caption{Energy and Fe moment (calculated and experimental) as a function of the $c/a$ ratio for stoichiometric FeTe (Fe$_{1.08}$Te). In the experimental data, $p$ is an implicit parameter}
\label{E_mom}
\end{figure}

\begin{figure*}[t!]
\centering
\includegraphics[width=1\linewidth]{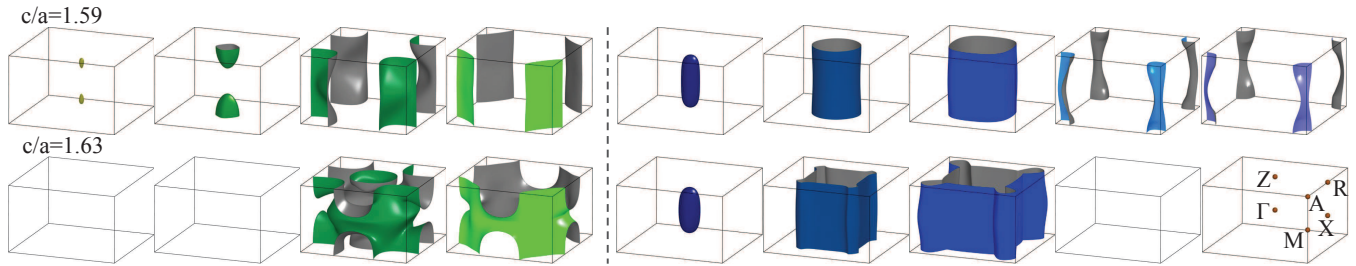}
\caption{Topological changes of the FS sheets for two closely spaced $c/a$ ratios in FeTe. The left four FS sheets correspond to the spin-up channel (green) and the right five ones to the spin-down channel (blue). Upon increasing the $c/a$ ratio by only 2.5\%, four FS sheets (two in each spin channel) disappear, illustrative of an ETT.}
\label{fermi}
\end{figure*}


For simplicity, we have limited our band-structure analysis to the tetragonal phase and stoichiometric composition. The energetics and magnetic moments are analyzed as a function of the $c/a$ ratio, whilst constraining the volume to that measured for 2.9~GPa (see previous paragraph). Firstly, we observe a salient feature in the total energy with increasing $c/a$ (see Fig.\,\ref{E_mom}): a non-monotonic flattening of the curve around $c/a \approx 1.6$, resulting in a shallow minimum. Upon further increase in $c/a$, the total energy then continues to decrease monotonically. Concomitantly, around $c/a \approx 1.6$, the Fe moment shows a sharp increase, strengthening from $0.4~\mu_{\mathrm{B}}$ to more than $1.5~\mu_{\mathrm{B}}$. Although the change in the total energy is subtle, the jump in the Fe moment is quite abrupt, and can be construed as being emblematic of the first-order character of the symmetry-conserving phase transition, consistent with our experiments.

Secondly, we monitored the change in the electronic structure of FeTe. Collected in Fig.\,\ref{fermi} are the FS's for two closely spaced $c/a$ ratios adjoining the jump in Fig.\,\ref{E_mom}. Upon increasing $c/a$ from 1.59 by 2.5\%, which is comparable to the experimentally observed increase at the symmetry-conserving transition, four FS sheets (two in each spin channel) rapidly reduce in volume and subsequently disappear. This is illustrative of the anticipated ETT, in the same vein as a Lifshitz transition \cite{Lifshitz_1960}. We note that we find a similar change in $c/a^*$ at the orthorhombic-incommensurate AFM to tetragonal-FM transition upon varying pressure.

Corresponding to the $c/a$ variation, a pronounced narrow peak in the density of states (see Supplemental Material \cite{suppl}) jumps from below the Fermi level to above, thereby shifting the Fermi energy away from the van Hove-like singularity. The loss of charge carriers due to the disappearing FS sheets is more than compensated by the remaining sheets which increase significantly in volume, consistent with the electrical resistivity becoming more metallic-like at higher pressures. Hence, the combined changes in the topology of the FS, concomitant with the depopulation of the narrow density of states peak, and the sharp increase of the Fe moment, demonstrate a Lifshitz-ETT in the spin-split bandstructure of Fe$_{1.08}$Te \cite{Lifshitz_1960,Rosner_2006}{\color{black}. Lifshitz already demonstrated that ETTs can drive
type-0 transitions \cite{Lifshitz_1960}, and it is known that electron-phonon coupling can exacerbate  anisotropic inharmonic response and instability of the lattice \cite{dagens79}.
However, the remarkable change of the $c/a$ ratio in Fe$_{1.08}$Te is unconventional and suggests that particular electron-phonon interactions exist in this system, so that the ETT results in an anisotropic lattice distortion. As we are observing a transition at finite temperature, its microscopic understanding should rely on the coupling of thermally smeared electronic states near singular Fermi-surface pieces to thermally excited acoustic phonons. Theoretically, such a strongly anisotropic effect in a $3d$-electron metallic system has not been anticipated, see review Ref.~\cite{blanter94}.}




{\color{black} Phenomenologically, the first-order thermal} PM-FM transition can be understood by considering the lattice strain, i.e. change of the $c/a$-ratio, as order-parameter, $\eta=e_{zz}$. It transforms as identity representation of the lattice space-group. Coupled to the magnetization $m$, the Landau potential for this transition reads \(f=\alpha\,m^2+\beta\,m^4 + A\,\eta^2+B\,\eta^3+C\,\eta^4+d\,\eta\,m^2\), where the last (attractive) term, $d<0$, encodes the unconventionally strong magneto-elastic coupling that relies on the ETT in the spin-split bandstructure. A first-order transition in the primary order parameter $\eta$ is prompted by the cubic term $B$, and can lead to a jump-like onset of magnetic order, if $|d|$ is comparable to $\alpha > 0$, as in similar systems with coupled ordering modes  \cite{holakowsky1973,Salje_2011}. {\color{black} This fundamental thermodynamic model describes all qualitative features and the strengths of the observed thermal transition in Fe$_{1.08}$Te, as a `type 0' process driven by the lattice instability $\eta$. In contrast, a magnetically driven transition would require to reach $\alpha < 0$ which entails only a weak response of the lattice (details in Supplemental Material \cite{suppl}).}

Another important structural aspect aiding the ETT induced FM state in Fe$_{1.08}$Te is the anisotropic {\it enhancement} of $c/a^{(*)}$ under pressure. An anomalous expansion of the lattice parameters under pressure has been observed in various members of the Fe-based superconductors, but they tend to reduce relatively quickly afterwards. They include the realization of a collapsed state, wherein the $c/a$ ratio is reduced considerably \cite{Kasinathan_2009,Kasinathan_2011,Chu_2009,Gati_2012,Guterding_2015}. The formation of interlayer bonds is seen as a driving force for the collapsed state, which in turn becomes nonmagnetic \cite{Kasinathan_2011}. In contrast, in Fe$_{1.08}$Te $c/a$ is {\it enhanced} in the FM state upon entering from the orthorhombic-incommensurate AFM phase or upon cooling from the tetragonal-PM one (Fig.\,\ref{phasediagram}). As seen in Fig.\,\ref{E_mom}, such an increase in $c/a$ is decisive in stabilizing the FM moment in Fe$_{1.08}$Te.

The formation of different magnetic structures due to an ETT is commonly observed in rare-earth containing metals (heavy-fermion systems) \cite{Chu_1970,Andrianov_2014}, wherein the localized moments of the rare-earth ions are interacting via the Ruderman-Kittel-Kasuya-Yosida (RKKY) mechanism \cite{Jensen_1991}. Small changes in the FS topology  of the conduction electrons which regulate the exchange mechanism can result in drastic changes of the magnetic properties. Dynamical mean-field theory studies on FeTe reported a mass enhancement of more than seven for the $3d_{xy}$ orbital, which is typical for heavy-fermion systems, and surmised that FeTe can promote an orbital-selective localization \cite{Yin_2010}. Another study based on semi-phenomenological models with coupled localized-itinerant moments in Fe$_{1+y}$Te, discussed the appearance of new magnetic  phases by the $y$-dependent RKKY part of the interaction \cite{Ducatman_2014}. In our work, projecting out the $3d$ orbital character, we notice the FS sheets to be comprised predominantly by the itinerant $3d_{xz/yz}$ and $3d_{x^{2}-y^{2}}$ orbitals (see Supplemental Material \cite{suppl}) while the $3d_{xy}$ orbital remains localized and away from the Fermi level. Consequently, the calculated rapid change in the FS topology, the sudden increase of Fe moment and the stabilization of the FM phase in Fe$_{1.08}$Te bear some similarity to heavy-fermion materials in that the moments on the localized $3d_{xy}$ orbital are mediated by the other itinerant electrons.  Thus, Fe$_{1.08}$Te presents a rare example of an ETT-induced FM transition under pressure in a $3d$ system, wherein only a handful of examples (mostly containing rare-earth ions) exist in the literature \cite{Kadomatsu_1986,Hauser_1994,Ishizuka_1997,Nakamura_2003,Tateiwa_2014}.

Our results identify a topological transition of the Fermi surface in the (spin-split) electronic band-structure of Fe$_{1+y}$Te that leads to a coupled instability of the lattice geometry with magnetic ordering and explains its unconventional magneto-structural 'type-0' transformation in the thermal energy range at a pressure of a few GPa. The results demonstrate the subtle competition between various electronic degrees of freedom and the coupling to (the geometry of) the lattice as the relevant parameter to explain the existence of the many different phases in this Fe-chalcogenide.



D.K. acknowledges funding by the DFG within FOR 1346. U.S., S.W., and S.R. acknowledge support by the DFG within SPP 1458.

\bibliography{FeTe}

\end{document}